\documentclass[showpacs,preprintnumbers,amsmath,amssymb]{revtex4}

\begin{document}

\preprint{}

\title{Error threshold in optimal coding, numerical criteria and classes of universalities
 for complexity}

\author{David B. Saakian$^{1,2}$}

\email{saakian@jerewan1.yerphi.am} \affiliation{$^1$Institute of
Physics, Academia Sinica, Nankang, Taipei 11529, Taiwan}
\affiliation{$^2$Yerevan Physics Institute,  Alikhanian Brothers
St. 2, Yerevan 375036, Armenia }

\date{\today}

\begin{abstract}

The  free energy of the Random Energy Model at the transition
point between
 ferromagnetic and spin
glass phases is calculated. At this point, equivalent to the
decoding error threshold in optimal codes, free energy has  finite
size corrections proportional to the square root of the number of
degrees. The response of the magnetization to the  ferromagnetic
couplings is maximal at the values of magnetization equal to half.
We give several criteria of complexity and define different
universality classes. According to our classification, at the
lowest class of complexity are random graph, Markov Models and
Hidden Markov Models. At the next level is Sherrington-Kirkpatrick
spin glass, connected with neuron-network models.  On a higher
level are critical theories, spin glass phase of Random Energy
Model, percolation, self organized criticality (SOC). The top
level class involves HOT design, error threshold in optimal
coding, language, and, maybe, financial market. Alive systems are
also related with the last class. A concept of anti-resonance is
suggested for the complex systems.

\end{abstract}

\maketitle

\section{Introduction}
{\bf Complexity}. The definition of statistical complexity is an
entirely open problem in statistical mechanics (see [1,2] for the
introduction of the problem and [3] for the recent discussion).
There are a lot of different definitions having sometimes common
context. A certain success was the discovery of an idea of
"schema", a highly compressed information, introduced by Gell-Mann
for complex adaptive (we assume, for all complex) systems.
    Some attempts, based mainly on entropy concepts, have been
    undertaken
    to define the concept of complexity. The approach [4-6], relevant
    for our investigation, is of special interest.
    A very interesting aspect of complex
     phenomenon is related to the
  edge of chaos (the border between
 chaotic and deterministic motion), the phase of  complex adaptive systems (CAS)
 [2],[7]. The concept of edge of chaos, independently suggested by P. Bak,
 S.A. Kauffman and C.Langton [8],  is not well defined
 quantitatively. However,
it is widely accepted this concept to be connected with the
sandpile [9,10]. This concept is of special importance  due to its
possible relation to the birth of life and evolution [7]. This
paper is devoted to relations between this phenomenon and some
aspects of information theory and optimal coding [11]. We assume
that the definition of a single (or best) complexity measure is a
subjective one, even with a reasonable constraint that complexity
should vanish for totally ordered or disordered motion  (see
dispute [12-13]). More strict is the definition of different
universality classes of complexity, which is presented in the
paper.
 We suggest several numerical criteria
for complex adaptive property. In practice we suggest to identify
the universality class of complexity from the experimental data
and to choose a model from the same class to describe the
phenomenon.

We assume the following picture of complex phenomenon. The
following hierarchy is presented: instead of microscopic motion of
molecules or spins we deal with the macroscopic thermodynamic
variables. Besides those, some new structures arose, sometimes
proportional to fractional degree of particles number. One can
understand qualitatively the complexity as a measure of new
structures. We are not going to scrutinize into the concrete
feature of those structures. We will just evaluate total measure
of structures on the basis of free energy expression, including
finite size corrections. The suggested complexity measures could
be applied for both pure systems, as well as to those defined via
disorder ensemble (as in spin-glasses). For interesting cases of
complex adaptive system,  a hierarchy in the definition of a
model, either a disorder ensemble (as in spin glasses), or scale
of the system (spatial or temporal) should be represented. The
structures themselves are derived from microscopic motions of
spins via order parameters. When those order parameter fluctuates,
they can be handled like the microscopic spins or molecules.
Therefore, in such cases, including the optimal coding of the
article, we can identify complex phenomenon as a situation with
changing reality or birth of new reality (the thermodynamic
reality is a mapping of molecular motions into few thermodynamic
variables). A good analogy is the weak interaction in high energy
physics on different scales. On the level low energies it can be
described through a simple picture by E. Fermi. However on the
level of 100GEV, when broken symmetry is restored, there is
absolutely new reality.

 Our
approach (to investigate just the ensemble averaged free energy
instead of logarithm of the number of different ground states in
spin-glasses [46]) is coherent with the idea of Jayenes [14]
saying that there is a single probability in physics, measurable
during observations, and there is no need to fracture it.

We assume that it is proper to define the criteria of complexity
via free energy  than via the entropy (see the discussion in
section three). In [15]  a criterion of complexity, functioning
from galaxies and stars to brains and society has been suggested:
the rate of free energy  density change. It is an interesting
approach, but we prefer to deal with dimensionless criteria. The
point that subdominant free energy describes the degree of system
complexity has already been recognized, at least, for 2d critical
systems.
 According to [16], the
subdominant term is proportional to conformal charge (effective
number of bosonic degrees of freedom). We just suggest to apply
this criterion for any system, as one of complexity criteria. For
the situation, when there is no explicit free energy (optimal
coding, sandpiles, etc.) one should try to find some equivalent
statistical mechanical formulation of the theory and investigate
free energy. In this work we give a first derivation of
subdominant free energy for the case, related to an optimal coding
and identify the universality class of error threshold. We provide
other criteria of complexity which have not been considered yet,
and the first list of universality classes.

{\bf Optimal coding}. Information processing should be certainly a
property of complex adaptive system. What can be clarified by
statistical physics? The connection of statistical physics to
information theory has been known from well known works of Jayenes
[14]. In 1989 N. Sourlas[17] found  a connection of Random Energy
Model (REM) of spin-glass by B. Derrida [18] with important branch
of Shannon information theory, i.e. optimal coding theory [11]. In
[19] I proved Sourlas idea for a principal case of finite velocity
codes. An important result has been derived by P. Rujan [20]
regarding to coding by statistical mechanics models at finite
temperatures. In a series of work [21-25] we derived the main
results of Shannon information theory using REM. Those results
have been repeated by alternative methods later (see review [26]).
Because we are going to consider models at the cross statistical
mechanics-optimal coding, some features of optimal coding  should
be briefly mentioned  [11].

Let us consider the transition of information, a sequence of $\pm
1$,  through a noisy channel to receiver. There exists an original
information which is a sequence of $+1$ and $-1$: $\epsilon_1\dots
\epsilon_N$.  If we send some letter through the channel, due to
the noise, the letters change their correct values, and
information is partially lost. Therefore, to recover further the
original message in a proper way it is needed to send originally
more information using a coding. The encoding is a mapping of the
initial message of length $N$ onto a sequence which has the length
$\alpha N$ ($\alpha >1$)
\begin{equation}
\label{e1} (\epsilon_1,\dots ,\epsilon_N)\to (f_1(\epsilon_1,\dots
,\epsilon_N),\dots f_{\alpha N}(\epsilon_1,\dots ,\epsilon_N)).
\end{equation}
 A noisy channel is represented as a mapping of the message by
 random letters $\eta_j$:
\begin{equation}
\label{e2} (f_1(\epsilon_1,\dots ,\epsilon_N),\dots f_{\alpha
N}(\epsilon_1,\dots ,\epsilon_N))\to (f_1(\epsilon_1,\dots
,\epsilon_N)\eta_1,\dots f_{\alpha N}(\epsilon_1,\dots
,\epsilon_N)\eta_{\alpha N}),
\end{equation}
where the noisy $\eta_j, 1\le j\le \alpha N$ are
 independent
random numbers with probability distribution
\begin{equation}
\label{e3}
P(\eta_j)=\frac{1+m_0}{2}\delta(\eta_i-1)+\frac{1-m_0}{2}\delta(\eta_i+1).
\end{equation}
The transmitter introduces additional information ($\alpha>1$),
and the receiver must extract useful information. These two
operations are called encoding and decoding. In general, coding is
a mapping of the initial message of length $N$ onto a sequence
which has the length $\alpha N$, $\alpha>1$. Thus, encoding is
done with $\alpha N$ functions $f_j=\pm 1$. The value
$\alpha^{-1}=R$, the "rate" of information transmission,
characterizes degree of redundancy. Decoding in general case is
called the procedure of extracting initial message out of noisy
sequence $(f_1\eta_1,\dots, f_{\alpha N}\eta_{\alpha N})$.

When is the errorless decoding  possible? We have a
Boltzmann-Gibbs-Shannon measure of information for a discrete
distribution $p_i$:
\begin{equation}
\label{e4} -\sum_ip_i \ln p_i.
\end{equation}
In original message any letter $\pm 1$ carries an information $\ln
2$. In case of noise by Eq. (\ref{e3}) any letter carries an
information
\begin{eqnarray}
\label{e5} \ln 2-h,\nonumber\\
h=-(\frac{1+m_0}{2}\ln \frac{1+m_0}{2}+\frac{1-m_0}{2}\ln
\frac{1-m_0}{2}).
\end{eqnarray}
In the last expression we extracted from the $\ln 2$ the entropy
$h$ of the distribution by Eq. (\ref{e3}).

The encoding is possible to work over in different ways. To
extract the original message without error it is reasonable to put
a constraint
\begin{equation}
\label{e6} \alpha N[\ln+\frac{1+m_0}{2}\ln
\frac{1+m_0}{2}+\frac{1-m_0}{2}\ln \frac{1-m_0}{2}]\ge N\ln 2.
\end{equation}
On the left, we have  information  of the received message. On the
right, we have an information to be  extracted. This is Shannon
fundamental theorem for errorless decoding. Only very special
coding schemes correspond to special case, when the last
expression transforms to equality. Such codes are optimal ones.
They are universal mathematical constructions, like critical
Hamiltonian in phase transitions.

{\bf Statistical mechanics for coding}. How could statistical
mechanics
 be applied for the optimal coding? To encode the original
sequence $\epsilon_1\dots \epsilon_N$, one constructs a
Hamiltonian $H(s)$, a function of $(s_1\dots s_N)$:
\begin{eqnarray}
\label{e7}
 -H(s_1\dots s_N)=f_1(s_1\dots s_N)f_1(\epsilon_1\dots
\epsilon_N)+\dots +f_{\alpha N}(s_1\dots s_N)f_{\alpha N
}(\epsilon_1\dots \epsilon_N)]\equiv
h_0(y_1,\dots y_{\alpha N}),\nonumber\\
y_j=
 f_j(s_1\dots s_N)f_j(\epsilon_1\dots
\epsilon_N).
\end{eqnarray}
 Here functions $f_j,1\le
j\le \alpha N$, are products of some $p$ spins, $f_j=s_{j_1}\dots
s_{j_p}$, and $H$ has a minimal value at $s_i=\eta_i$.  The
influence of noise is very simple: every term (word) in (\ref{e7})
is multiplied by a noise, and instead of pure Hamiltonian $H(s)$
we have a noisy one
\begin{eqnarray}
\label{e8}
 -H(s,\eta)=h_0(y'_1\dots y'_{\alpha N}),\nonumber\\
y'_j=
 f_j(s_1\dots s_N)f_j(\epsilon_1\dots
\epsilon_N)\eta_j.
\end{eqnarray}
 To find the minimum of the Hamiltonian, one could consider a
statistical mechanics of the spin system with the Hamiltonian $H$
at very low temperatures,
\begin{equation}
\label{e9}
 Z=\sum_{s_i=\pm 1}e^{-\beta H(s,\eta)},
\end{equation}
where $H(s,\eta)\equiv H(s_1\dots s_N,\eta_1\dots
\eta_N)$,$\beta\to \infty$. Without noise ($\eta_j=1$) one can
calculate the configuration $(s_1\dots s_N)$ giving the main
contribution to $Z$ at $\beta\to \infty$. We have the following
expression for the mean magnetization:
\begin{equation}
\label{e10} <s_i>=\epsilon_i.
\end{equation}
It has been proved in [21] that Eq. (\ref{e10}) is correct also
for the non-zero noise below the Shannon error threshold.

In Shannon information theory one considers transmission of a
message to a receiver. The influence of the noise corresponds to
simple product of coding words $f_j(\epsilon_1\dots \epsilon_N)$
by a noisy letter $\eta_j$ (both are accepting the values $\pm
1$).

{\bf Other versions of error threshold in statistical mechanics}.
What generalizations of the considered scheme are possible to
accept? Instead of Eq. (\ref{e9}) one can consider a partition
with the quantum noise:
\begin{equation}
\label{e11}
 Z=Tr \exp\{-\beta [H(\sigma^z_1\dots
 \sigma^z_N)+\mu\sum_{j=1}^N\sigma^x_i]\},
\end{equation}
and
\begin{equation}
\label{e12}
 Z=Tr \exp\{-\beta [e^{\gamma(1-\sum_{j=1}^N\sigma^x_i/N)}H(\sigma^z_1\dots
 \sigma^z_N)]\},
\end{equation}
where $H$ is a mean field like Hamiltonian like to
\begin{equation}
\label{e13}
 H(\sigma^z_1\dots \sigma^z_N)\equiv H_0(\sum_i\sigma^z_i),
\end{equation}
having minimum at configuration $s_i=1,1\le i\le N$. The
successful information transmission is connected with the phase,
where $<\sigma^z_i>\equiv m_i>0$ (there is a non-zero longitudinal
magnetization),
 in Eq.(\ref{e11})
quantum noise is additive, in Eq. (\ref{e12}) it is a
multiplicative one. Eqs. (\ref{e11},\ref{e12}) are connected to
the evolution models \cite{ei71, bbw97,sh04,sh04a}, when genetical
information is transmitted to future generations. It is
interesting that Eigen derived correct error threshold  in its
model [27] just from informational theoretical arguments long
before the Sourlas work about connection of statistical mechanics
with information theory. Eigen model has been exactly solved only
recently \cite{sh04},   Eigen's  formula for error threshold was
confirmed.

Our purpose is to connect the complex adaptive phase with the
neighborhood of error threshold (\ref{e6}). We will consider the
border between ferromagnetic and spin glass phases in Random
Energy Model, investigating its statistical mechanics by Eq.
(\ref{e9}). We will not consider the informational-theoretical
aspects of the problem any more-the subject is well discussed in
[25].  In section 2 we will derive the finite size correction to
free energy and investigate the dependence of magnetization from
the bulk ferromagnetic coupling. In section 3 we will give a
definition of complex adaptive property and define different
universality classes. In section 4 we will suggest another concept
of complex adaptive systems, i.e. the possibility of
anti-resonance. In conclusion we will briefly discuss our results
and general aspects of complex adaptive systems.

\section{Random Energy Model}
{\bf Energy configuration formulation.} To investigate the complex
phenomenon we consider an equilibrium statistical physics
situation
 similar to  edge of chaos point i.e.  the border between ferromagnetic and  spin-glass (SG)
phases in Random Energy Model (REM). The finite size corrections
of free energy will be calculated later on. In REM  N spins
$s_i=\pm 1$ interact through $(^p_N)\equiv\frac{N!}{p!(N-p)!},p\to
\infty$ couplings with the Hamiltonian [18,21]
\begin{equation}
\label{e14} H=-\sum_{1\le i_1..\le i_p\le N}[j^0_{i_1..i_p}+
j_{i_1..i_p}]s_{i_1}.. s_{i_p}.
\end{equation}
Here $ j^0_{i_1..i_p}$ are ferromagnetic couplings
\begin{equation}
\label{e15} j^0_{i_1..i_p}=\frac{J_0N}{(^p_N)},
\end{equation}
 and for quenched disorder $j_{i_1..i_p}$ we have a distribution
\begin{equation}
\label{e16}
\rho_0(j_{i_1..i_p})=\frac{1}{\sqrt{\pi}}\sqrt{\frac{(^p_N)}{N}}
\exp\{-j_{i_1..i_p}^2\frac {(^p_N)}{N}\}.
\end{equation}
We see that there are ferromagnetic and random couplings, and
$J_0$ defines the ferromagnetic degree.

In our spin model there are $2^N$ different energy configurations.
It has been found by B. Derrida that for large values of p there
is a factorization for energy level distribution. For
$\alpha\ne\beta$ [18]
\begin{equation}
\label{e17}
\rho(E_{\alpha},E_{\beta})=\rho(E_{\alpha})\rho(E_{\beta}).
\end{equation}
 For the 1-st configuration with $s_i=1$ [21]:
\begin{equation}
\label{e18} \rho_1(E_1)=\frac{1}{\sqrt{\pi
N}}\exp[-(E_1+J_0N)^2/N],
\end{equation}
and for other $2^N-1$ levels [18]
\begin{equation}
\label{e19} \rho(E)=\frac{1}{\sqrt{\pi N}}\exp(-E^2/N).
\end{equation}
REM has two equivalent  definitions: via energy configuration Eqs.
(\ref{e18},\ref{e19}) and via spin Hamiltonian version Eq. (14).
It is possible to solve REM through ordinary spin glass approach,
as well as using the factorization property Eq. (17). According to
energy configuration approach,  we perform averaging via energy
level distribution
 (instead of random couplings in usual case of disordered systems):
\begin{equation}
\label{e20} <\ln Z>\equiv <\ln\sum_{\alpha}\exp{(-\beta
E_{\alpha})}>_E.
\end{equation}
Here $\beta$ is an inverse temperature. It is possible to derive
[21] that at high enough values of $J_0>\sqrt{\ln 2}$, (see Eq.
(\ref{e27})), at low temperatures the system
 is in ferromagnetic phase with magnetization
\begin{equation}
\label{e21} m_i=1.
\end{equation}

 Using the trick [18],
\begin{equation}
\label{e22}<\ln Z> =\Gamma'(1)+\int_{-\infty}^{\infty}\ln t
\frac{d <\exp[-tZ]>}{d t} {\it d}t.
\end{equation}
one can factorize the integration via different energy levels
$E_{\alpha}$. The average is over energy distributions Eqs.
(18),(19). It is enough only to calculate for $<\exp[-te^{\beta
E_i}]>$ for the single level. We consider
\begin{eqnarray}
\label{e23} f(u)\equiv \frac{1}{\sqrt{\pi}
}\int_{-\infty}^{\infty} \exp[-y^2-e^u \exp(-\lambda y)]dy,
\end{eqnarray}
where $\lambda=\beta\sqrt{N}$ and $<\exp[-te^{\beta E_i}]>=f(\ln t
)$. We can further derive for the $<e^{-tZ}>\equiv
<\exp[-te^{\sum_{\alpha=1}^ME_{\alpha}}]>$
\begin{eqnarray}
\label{e24}
 \Psi(u)=[f(u+u_f)f(u)^M],
\end{eqnarray}
where $u=\ln t, u_f=J_0N\beta, M=2^N-1$. Now Eq. (\ref{e22}) gives
\begin{eqnarray}
\label{e25} <\ln Z>=\Gamma'(1)+\int_{-\infty}^{\infty}u\frac{d
\Psi(u)}{d u} {\it d}u.
\end{eqnarray}
 $f(u)$ is a monotonic function. With exponential accuracy it equals $1$
below $0$, then becomes $0$ above it. We need four asymptotic
regimes [18,21]:
\begin{eqnarray}
\label{e26} f(u)&\approx&\frac{1}{\sqrt{\pi}\lambda}\Gamma
(\frac{2u}{\lambda^2}) e^{-\frac{u^2}{\lambda^2}},
 \lambda \ll u \nonumber\\
f(u)&\approx&\frac{1}{\sqrt{\pi}}\int_{u/\lambda}^{\infty}dx
e^{-x^2},
  \mid u \mid \ll \lambda^2\nonumber\\
f(u)&\approx& 1-\frac{1}{\sqrt{\pi}\lambda}\Gamma
(-\frac{2u}{\lambda^2}) e^{-\frac{u^2}{\lambda^2}},
-\frac{\lambda^2}{2}<u\ll-\lambda\nonumber\\
f(u)&\approx& 1-e^{u+\frac{\lambda^2}{4}},
-\lambda^2<u<-\frac{\lambda^2}{2}.
\end{eqnarray}
We are interested in those asymptotic for $u\sim N$ or $u\sim
\sqrt{N}$ and $\lambda\gg 1$.  As the $f(u+u_f)f(u)^M$ is like to
step function, its derivative is like to $\delta$ function with
center at some $-u_0$. The vicinity of $-u_0$ contributes mainly
to the integral in (\ref{e25}) (bulk value is equal to $u_0$).
Ferromagnetic phase appears, when the $-u_f$ (the center of
function $f(u+u_f)$ is lefter, than $-\sqrt{N}\lambda\ln 2  ($ the
center of $f(u)^M$). The FM-SG border corresponds to:
\begin{equation}
\label{e27} J_0=\sqrt{\ln 2},\qquad \infty>\beta>\sqrt{\ln 2}.
\end{equation}
When there is only the first level with distribution (\ref{e18}),
$<\ln Z>\equiv -\beta <E_1>=u_f\equiv J_0N\beta$. For that case
$\Psi=f(u+u_f)$. Therefore Eq. (\ref{e25}) gives:
\begin{equation}
\label{e28} \Gamma'(1)+\int_{-\infty}^{\infty}ud[f(u+u_f)]=u_f.
\end{equation}
Using the last identity, we transform Eq. (\ref{e25}) into
\begin{eqnarray}
\label{e29} <\ln Z>= \Gamma'(1)+\int_{-\infty}^{\infty}ud\Psi(u)
=u_f-\int_{-\infty}^{\infty}ud\Psi_1(u)
=u_f+\int_{-\infty}^{\infty}\Psi_1(u)du,
\end{eqnarray}
where $\Psi_1(u)=f(u+u_f)[1-f(u)^M]du$.

{\bf Exact border of ferromagnetic and spin glass phases.} Let us
first consider the exact border of two phases $J_0=\sqrt{\ln 2}$.
$\Psi_1(u)$ is a product of two monotonic functions, decreasing
(one-to the left, another- to the right)  from the point $u=-u_f$.
We define an auxiliary function $F(u)$ by differential equation:
\begin{equation}
\label{e30} F^{'}(u)=f(u+u_f).
\end{equation}
Using the second equation in (\ref{e26}), we derive for $  \mid u
\mid \ll \lambda^2$:
\begin{eqnarray}
\label{31} F(u-u_f)=\int_{0}^{u/\lambda}dx
\frac{\lambda}{\sqrt{\pi}}\int_{x}^{\infty}e^{ -y^2}dy.
\end{eqnarray}
Let us denote $\Psi_2(u)=(1-f(u)^{M})$ and perform integration by
parts in Eq. (\ref{e29}):
\begin{eqnarray}
\label{e32} <\ln
Z>&=&u_f+\int_{-\infty}^{\infty}F^{'}(u)\Psi_2(u)du=u_f+
F(\infty)\Psi_2(\infty)-F(-\infty)\Psi_2(-\infty)
-\int_{-\infty}^{\infty}F(u)\Psi^{'}_2(u)du\nonumber\\&=&u_f+
[F(\infty)-F(-u_f)]
-F^{'}(-u_f)\int_{-\infty}^{\infty}(u+u_f)\Psi^{'}_2(u)du.
\end{eqnarray}
We have truncated expansion in degrees of $u+u_f$ as
$\Psi^{'}_2(u)$ is similar to  $\delta$ function near the $-u_f$.
We used $\Psi_2(\infty)=1,\Psi_2(-\infty)=0$ and
$F(-u_f)\int_{-\infty}^{\infty}\Psi^{'}_2(u)du=F(-u_f)$. Then Eq.
(\ref{e32}) gives:
\begin{equation}
\label{e33} <\ln Z>-u_f\approx \frac{\beta
\sqrt{N}}{\sqrt{\pi}}\int_{0}^{\infty}dx\int_{x}^ {\infty}\exp
(-y^2)dy\sim N^{\frac{1}{2}}.
\end{equation}
Eq. (\ref{e33}) is one of the main results of our investigation.
It is obvious that besides the bulk term in free energy asymptotic
there is a subdominant term proportional to square root of the
number of degrees. The importance of the subdominant term in
entropy has been underlined in [4], and has been well analyzed in
[5]. They suggested to identify different universality classes of
complex phenomena by the subdominant terms of entropy. In 1-d
spin-glass model with long-range interaction, $<J_{ij}^2>\sim
1/(i-j)^2$, they derived Eq. (\ref{e33}) for the entropy. In [5]
another object with a similar subdominant entropy has been
mentioned,i.e. language [30].

{\bf Small deviation from the border of two phases.} Consider
small deviation from Eq. (\ref{e26}) (scaling is reasonable, as we
see in Eq. (\ref{e35})):
\begin{equation}
\label{e34} J_0=\sqrt{\ln 2}+\frac{j_0}{\sqrt{N}}.
\end{equation}
Now the finite size correction are less than in Eq.(33) and
decrease exponentially at large values of $j_0$:
\begin{eqnarray}
\label{e35} <\ln Z>-(\sqrt{\ln 2}+\frac{j_0}{\sqrt{N}})N \sim
\beta \sqrt{N}\exp[-j_0^2].
\end{eqnarray}

Now calculate the magnetization. We define:
\begin{equation}
\label{e36}
\begin{array}{l}
m=<\frac{\exp(-\beta E_1)}{\sum_{\alpha}\exp(-\beta E_{\alpha})}>.
\end{array}
\end{equation}

Using  the identity $\frac{1}{Z}=\int_0^{\infty}dte^{-t Z}$
 we derive for $m$
\begin{equation}
\label{e37}
 m=-\int_0^{\infty}d t\frac{d}{d t}f(u+u_f)f(u)^M
=1-\int_{-\infty}^ {\infty}duf(u+u_f)\frac{d}{d u}f^M(u),
\end{equation}
where $u=\ln t$. Using  second equation in Eq. (26) we derive
\begin{equation}
\label{e38} m=\frac{d<\ln Z>}{\beta \sqrt{N}d j_0} =
\frac{1}{\sqrt{\pi}}\int_{-j_0}^{\infty}\exp [-y^2]dy.
\end{equation}
And for its differential:
\begin{equation}
\label{e39} \frac{d m}{d j_0}=\frac{1}{\sqrt{\pi}}\exp [-j_0^2].
\end{equation}
The last expression could be represented also as
\begin{equation}
\label{e40}\frac{1}{\beta \sqrt{N}} \frac{d^2<\ln Z>}{d^2 j_0} =
\frac{1}{\sqrt{\pi}}\exp [-j_0^2].
\end{equation}
 Thus,  at the exact border
SG-FM ($j_0=0$)  the dependence of magnetization  from the
external (ordered)
 parameter is maximal (maximum instability principle). This is likely a characteristic
  property of every  complex adaptive
 system (CAS).
 One has an  ordered external parameter to manage the system as well as
 random parameters (the choice of "ordered" and "random" could be subjective).
  There is an emergent (essentially collective) property. If the interaction with an environment is defined via the
emergent property, then CAS drifts to the
 maximal instability point with maximal dependence of this emergent property from the
 ordered parameter.

One can consider the $\frac{d m}{d j_0}$ as some degree of
complexity. A close characteristic is the second derivative of
free energy via ordered coupling, Eq. (\ref{e40}). In our case,
they coincide. However, there are possibly more complicated
situations, when they are different and both should be used.

Let us calculate the moments of $P_{\alpha}\equiv
\frac{\exp(-\beta E_{\alpha})}{ \sum_{\delta}\exp(-\beta
E_{\delta})}$. Using the identity
\begin{eqnarray}
\label{e41} \frac{1}{\sqrt{\pi} }\int_{-\infty}^{\infty}
\exp[-y^2-n\lambda y-e^{u -\lambda y}]dy=\frac{d^n f_0(t)}{dt^n},\nonumber\\
f_0(t)\equiv f(\ln t).
\end{eqnarray}
We have:
\begin{eqnarray}
\label{e42} <P_1^2>&=&\int_0^{\infty}tdt
f_0(t)^{M-1}\frac{d^2f_0(te^{J_0N\beta})}
{dt^2}=\int_{-j_0}^{\infty}\frac{e^{-x^2}}{\sqrt{\pi}}dx,\nonumber\\
<P_{\alpha}^2>&=&\int_0^{\infty}tdt f_0(t)^{M-2}\frac{d^2f_0(t)}
{dt^2}f_0(te^{J_0N\beta}),\nonumber\\
<P_{\alpha}P_{\gamma}>&=&\int_0^{\infty}tdt f_0(te^{J_0N\beta})f_0(t)^{M-3}(\frac{df_0(t)}{dt})^2,\nonumber\\
\sum_{\alpha,\gamma>1}<P_{\alpha}P_{\gamma}>&=&1-
\frac{1}{\sqrt{\pi}}\int_{-j_0}^{\infty}\exp [-x^2]dx.
\end{eqnarray}
For the $T<T_c\equiv 2\sqrt{\ln 2}$:
 \begin{eqnarray}
\label{e43} \sum_{\alpha>1}<P_{\alpha}^2>&=&[1-
\frac{1}{\sqrt{\pi}}\int_{-j_0}^{\infty}\exp [-x^2]dx][1-\frac{T}{T_c}],\nonumber\\
\sum_{\alpha,\gamma>1,\alpha\ne
\gamma}<P_{\alpha}P_{\gamma}>&=&[1-
\frac{1}{\sqrt{\pi}}\int_{-j_0}^{\infty}\exp
[-x^2]dx]\frac{T}{T_c}.
\end{eqnarray}
Define
 \begin{eqnarray}
\label{e44}
 C=
<P_1^2>\sum_{\alpha>1}<P_{\alpha}^2>.
\end{eqnarray}
$C$ takes the maximal value at the critical point $j_0=0$
 \begin{eqnarray}
\label{e45} C=\frac{T_c-T}{2T_c}.
\end{eqnarray}
 For large
 $j_0$ we have that  $C$ decreases exponentially:
 \begin{eqnarray}
\label{e46} C \sim \exp [-j_0^2].
\end{eqnarray}

 The more detailed investigation of $j_0=0$
case states that
 \begin{eqnarray}
\label{e47} <P_1>=\frac{1}{2},<P_1^2>=\frac{1}{2}.
\end{eqnarray}
We see, that $P_1=0,1$  with probabilities $1/2$.

We can define $C$ as edge of chaos parameter. At the exact error
threshold border it has a maximal value, equal  $1/2$ at zero
temperature,i.e. the probabilities of ordered and random motions
are equal.  $C$ decreases exponentially outside the region. What
is the advantage of our choice Eq. (\ref{e44}) over another one,
$<p_1>\sum_{\alpha>1}<P_{\alpha}>$? Eq. (44) distinguishes the
$\beta\to\infty$ as the most optimal situation, and the last
choice fails.

To define $C$, we have actually used the Tsallis entropy  at $q=2$
[32]
 \begin{eqnarray}
\label{e48} I_q=-\frac{[\sum_rp_r^q-1]}{q-1}.
\end{eqnarray}
In [3] M. Gell-Mann and S. Lloyd assumed the connection of $I_q$
with the edge of chaos systems.

\section{Definition of complex adaptive property and universality classes}

{\bf Definition of complexity}. Free energy is the fundamental
object in statistical mechanics. The bulk free energy is
proportional to the number of particles (spins). It is well known
that in case of some defects on geometrical manifolds (lines,
surfaces), besides the bulk term in the asymptote expression of
free energy, there are subdominant terms proportional to some
roots of $N$. Thus, the subdominant term in free energy could be
identified with existence of some structures (much more involved
than simple geometrical defects) in the system. In our case of
REM, the formulation of the model was homogeneous in the space,
but we got a square root subdominant term. In complex system we
assume the following hierarchy: bulk motion and some structures
above it. The subdominant free energy is related to the
structures. If we are interested just in structure, we can ignore
 bulk free energy (an analogy in the physics of surface: to
investigate the surface free energy we certainly miss the bulk
energy). Therefore:

{\bf A.} We define the complexity as subdominant free energy.\\ We
have seen that in case of error threshold via  REM it scales as a
square root of number of spins. We assume that it is the most
important class of complex phenomena, connected with alive
systems. In complexity phase intermediate scale free energy (or
entropy, or Kolmogorov complexity) becomes strong, and the
subdominant term scales as a square root with the number of
degrees. Eqs. (\ref{e33}),(\ref{e35}).

 What do we mean by intermediate scale? There
is a minimal scale (ultraviolet cutoff) and maximal scale
(infrared one). The intermediate scale is just their geometric
average.
 In [33] has been investigated
the statistics of the heartbeats. They found that healthy people
can be differentiated by coarse-grained entropy at the
intermediate scale, which is
 coherent with appearance of middle scale free energy in our case.
 Therefore the situation is coherent
with the criterium A. The complexity in our definition is free
energy on a higher hierarchy level (connected with the
structures). One should remember that free energy itself is a
second level on hierarchy. The energy is on the ground level is .
Due to thermodynamic motion, only its smaller part is manageable
on macroscopic level (only free energy could be extracted as a
mechanical work while changing the global parameters). Therefore,
complexity is a level on a hierarchy of the following modalities:
energy,free energy and subdominant free energy ones. Every higher
level is more universal. It is explicit in quantum field theory
approach to critical phenomenon [16]. Different renormalization
schemes can give different bulk free energies, but the same
logarithmic subdominant one.  Thus we observe a hierarchy of
modalities (non-categorical statement about reality, see [34]). In
principle, the hierarchy could be continued, and at some level the
life could appear. Our view (rather statistical mechanical, than
mathematical) is close to the one by M. Gell-Mann and S. Lloyd in
[2], defining system complexity as "length of highly compressed
description of its regularities".

Due to above  mentioned hierarchy, the identification of
complexity with a subdominant free energy is more universal, than
the entropy approach of [4,5]. Sometimes the existence of
structure could be identified in entropy or Kolmogorov complexity
subdominant terms as well. In our case the free energy (sic!)
reveals a huge subdominant term, but not the entropy.

We assume that other features of our toy model are characteristic
for complex adaptive systems:

{\bf B.} There is an emergent property, maximally unstable under
the change of ordered external parameters, Eq.(\ref{e39}).
Sometimes it can be characterized as a second derivative of free
energy via ordered parameter.

{\bf C.} The probability of ordered and disordered motions should
be at the same level (like to Eq. (45),(47)).

{\bf D.} The complex adaptive properties could be exponentially
damped
   in case of even small deviation of ordered parameter.

Let us discuss different complex systems, defining universality
classes.

{\bf Critical theories}. We assume that subdominant term of free
energy
 describes the number of
 real parameters of the system. In [5] has been considered a
 learning process for a
model with finite $K$ parameters and a logarithmic subdominant
term, proportional to $K$ has been found. For  2d critical
theories we can take  either total effective number of bosonic
degrees (conformal charge $c$), or the number of primary fields as
a number of parameters. According to [16], nature has chosen the
first one, and subdominant term of free energy is proportional to
the conformal charge and to the logarithm of the degrees of
freedom.

We see that complex phenomena analyzed in previous sections
correspond to another class of universality that the models in
[16]. In critical theories [16] magnetization disappears at the
transition point (contrary to error threshold case). Therefore, we
admit that complex adaptive system, while having some scaling (fat
tails in
 markets), could not be described by  critical
theories.

In 2d percolation indices could be described by conformal field
theory. Therefore, the percolation belongs to the class of
critical theories. In our classification such situation is as
complex as the class [16].

In spin glass model of REM B. Derrida found a logarithmic
subdominant free energy. Therefore, the model belongs to the class
of [16].

 {\bf Financial markets}. One can apply our criteria to
financial markets [35].  To analyze the financial time series
$y(t)$  (US dollar-German mark exchange rate)the statistics of
price increment $y(t+\tau)-y(t)$ has been considered and
probability density function (pdf) $p(x,\tau)$ has been
constructed from the empiric data. A Fokker-Planck equation, where
the role of time plays $\ln \tau$, has been derived  for the last
distribution. We see a diffusion in the scale $\ln \tau$ as well
as a drift. In [36] the ratio $R$ of ordered motion of $y$ and the
diffusion has been calculated. It is the tail exponent of the $y$:
$P(y)\sim \delta y^{-(1+\mu)}$ [36]. In practice, $R=\mu\sim 3-5$.
In the situation, when the approach of [35-36] is correct, the
more complex situation corresponds to the smaller values of $\mu$.
In case of error threshold model, considered in this paper, the
subdominant term is larger in the region $R\sim1 $. Outside, it
decreases exponentially like the one in Eq. (\ref{e46}).

For the markets something like the property can be also observed.
 There are fundamentalist traders who act in a
deterministic way and the noisy ones [37]. In our model, they are
similar to ordered and random couplings. In case of B, the
fundamentalists' number is chosen to have a maximal influence to
the market global characteristics.  In usual thermodynamics we
have a fundamental notion of temperature, and the equilibrium is
possible only when temperature of different subsystems is the
same. Now  an edge of chaos parameter for the complex adaptive
systems is introduced. It is reasonable to assume that in stable
state it should be the same for different parts of the market (for
example, for the traders and stocks). In this way it could be
possible to predict future catastrophes. One can identify the edge
of chaos parameters also considering a correlation matrix of
different stocks. According to the above mentioned  data, there
are both deterministic and stochastic parts. It is very important
to identify the subdominant term in entropy considering block
entropies of financial data.

{\bf Highly Optimized Tolerance (HOT) design}. It is the last
crucial achievement of complex system theory, related to the
robustness of engineering design [38-39].  In the simplest case
one considers fire forest model on 2-d lattice. There are trees at
any site of lattice, and there is a known probability of sparks.
As a tree is fired, its nearest neighbors are also fired. One
constructs firebreaks (sites without trees) to limit the size of
the event (total number of fired trees). The goal is to construct
a robust scheme against fire propagation for the given spark
probability, using a minimal area of firebreaks. Scaling laws for
the distribution of fire events has been found.
 The situation highly
resembles error threshold case. Actually, in [39] the connection
of HOT design with source coding has been directly stated.  In
error threshold there is also
 scaling for the mean magnetization $m=1/2+c/\sqrt{N}$ [22].
It has been assumed in [38,39] that SOC and HOT design are
different classes of universality. We can adduce another argument.
 In sandpile there is an analog of free energy, the number of recurrent
states of the sandpile process. It is the number of spanning trees
of free-fermions model. Therefore sandpile belongs to the class of
universality [16] with central charge $c=-2$ [10]. We have used an
important principle: the class of universality of the complex
system should be the same in all of its representations. A very
interesting feature of HOT design is that it gives a robustness
against the originally given distribution of the noise. The
robustness is very fragile: there is a large probability for the
total crush (great fire) in case of the change of the original
conditions. This resembles property D in the definition of complex
adaptive phenomenon. In [40] has been suggested a constraint
optimization with limited deviation (COLD) design  to avoid large
probability of total crush. They also mentioned  the first known
example of HOT design like situation. It is the classic problem of
gambler's ruin: the optimizing total return leads to ruin with
probability one \cite{co91}.  For a very complicated complex
system with many hierarchies, the full optimization states a
single simple principle for a management over the system, as in
such case the essence of different hierarchies should be the same.
Only the absolute optimization allows a full transformation of the
content from one hierarchy level to another. This crucial feature
has been lost in COLD case. I think that the choice of COLD can be
successful only for not too much complicated systems. In the next
section we will discuss a close concept of anti-resonance for the
complex adaptive systems, exploring the property D in our
approach.

{\bf Markov models and random networks.} There are a lot of
applications of Markov models in complex systems. Especially
important are applications in bioinformatics \cite{ko01}. There is
some biological language in DNA and proteins, and hidden Markov
models (the transition between states of the system is observed in
a probabilistic way) have been applied to model this language. One
can investigate the block entropies $S(N)$ for the words with $N$
letters in the stream of data and define the subdominant term.
Such investigation has been thoroughly done in [6]. At large $N$,
in case of classic order $R$ Markov process $S(N)$ gets an exact
linear asymptote at $N>R$. For the case of hidden Markov model  a
subdominant entropy, decreasing exponentially with $N$ has been
found. This is very important moment. Those models, being very
useful, don't share the class of universality of alive systems,
which we assume as corresponding to the subdominant term $\sim
\sqrt{N}$.

Networks are very popular in complex system research. How these
geometrical objects could be classified into universality classes?
In [43] has been introduced a statistical mechanics approach to
describe the properties of a graph ensemble. The mean
characteristics of the graph has been fixed, while maximizing the
entropy of the ensemble. Now the number of pairs of vertices plays
the role of number of degrees of freedom. The free energy can be
defined.  For the case of random graph there is no finite size
correction in free energy expression. Therefore, random graph
corresponds to the Markov model class of complexity. Unfortunately
I don't see a way to enlarge the method of [43] to scale-free
networks.

  {\bf Virus evolution near
the error threshold.} The evolution of the majority of viruses
(RNA genome viruses) is described well by Eigen model [27].  This
brilliant model gives a simple and complete version of Darwin
evolution theory. Information is represented here as a chain of
spins taking $\lambda=2$ or $\lambda=4$ values. There are
$\lambda^N$ different configurations with corresponding
probabilities $p_i,1\le i\le \lambda^N$. At any moment, the virus
is giving offsprings with some rate specific for his genome
(fitness). Offsprings randomly change their mother genome to other
ones (mutations). When the majority of individuals has a genome
near one configuration ("wild" one), then genetic information is
successfully transferred to future generations. Otherwise, there
is a flat distribution of individuals in the genome space. It is
interesting that the virus evolution is near the error threshold.
In "quasispecise theory" [44] (virus population with a
distribution like a cloud around some "wild" genome configuration)
there are equivalents of energy,i.e. fitness, free energy,i.e.
mean fitness for the whole system (for one configuration a product
of fitness and errorless copying probability). All of these
(selective abilities) can be
 derived in this model just as a
consequence of Eigen equations. During the evolution, population
is located mainly in a genomes with high selection ability.
Considering the evolution in dynamic environments, it is possible
to define a new kind of selective ability, like higher form of
free energy (complexity?). Such approach to define a complexity is
 quite objective one. We assume that it is possible to calculate
analytically also the ground state entropy (including the
subdominant one) and define the complexity by [4-5].

It could be possible to investigate some aspects of optimal
coding, impossible to do in alternative way. Choosing as REM's
Hamiltonian like fitness function, we can get an analytical
dynamics for optimal coding (the work is in progress). Thus,
rigorous investigation of informational theoretical (complexity)
aspects of evolution models could be very fruitful for both
disciplines.

Virus evolution is often referred to as a typical example of
complex adaptive system. Another famous example is an immune
system. Statistical mechanics has been successfully applied to
this case [45]. I don't see  a direct analogy with the error
threshold phenomena here. But one should definitely choose model
 from high complexity class.

{\bf Sherrington-Kirkpatrick model.} Usually one defines the
logarithm of different ground states [47] (solutions of
Thouless-Anderson-Palmer equations) as a complexity. It is a
reasonable characteristics to be investigated (although  very
complicated one). I think  that to identify the universality class
of the model it is enough to calculate finite size corrections of
free energy or energy. Such calculations have been done for a
Sherrington-Kirkpatrick model [48]. The subdominant energy scales
as $N^{1/3}$. Therefore, it is a new class of complexity. For
different spin-glasses other subdominant term scalings are
possible as well, and finite dimensional spin-glasses are likely
to have another universality class. We have mentioned the
Sherrington-Kirkpatrick model [46] , because it is connected with
neuron-networks.

\section{Anti-resonance in complex systems}

{\bf Complex resonance}. The concept of resonance is probably the
most noticable phenomenon in nature, culture and science.  The
close notion of synchronization in complex system is becoming more
and more popular [49].  We are going to analyze the idea of
resonance in complex systems, to look for a possibility of, in
some sense, inverse situation with an exponential damping of
motion (anti-resonance). We suppose that this notion will
compliment our view to complex systems in previous section.

Originally, the simplest resonance situation has been investigated
in mechanics of classical deterministic system with some resonance
frequency, driven by external harmonic force. When two frequencies
coincide, the reaction of the system to external force increases
drastically. Even in this simple case we can observe two features
of phenomenon. Frequency is an essence of motion, and there is a
sharp peak in the ratio output-force.

The next step was parametric resonance in classical mechanics.
There is a hierarchy here. We observe a motion at given values of
parameters, and the resonance frequency depends on the value of
external parameters. If one changes the external parameter with
the same frequency, as the frequency of the pendulum, there
appears famous parametric resonance-the flow of energy from the
higher level of hierarchy to the lower level one. Let us
generalize this situation to other complex systems to define
complex resonance.

If there is a hierarchy in the system, and states at different
 hierarchic levels have some essence (comparable logically with
 each other), the generalized resonance happens, when
 these essences coincide.
What about the essence of the state? In classical mechanics, there
is only one real number characterizing total state, i.e.
frequency. In general one should look for
 other total parameters of the system. In modern physics
 these are the following: temperature in statistical mechanics, replica system
  breaking scheme in spin-glasses (edge of chaos parameter), and the wave function phase in
   quantum mechanics.

The next famous example of such (generalized parametric resonance)
situation is related to the Nishimori line in disordered systems
[50,51]. A hierarchy (quenched disorder) is present here.
Sometimes
 it is possible to introduce some formal temperature to describe
 this disorder.  If two temperatures (real
 one for the spins and the formal one for the quenched disorder)
  coincide, the system reveals some interesting properties becoming
  maximally analytic in some sense.

So we can define a hierarchy for the resonance. In the trivial
case, the system is not hierarchic, it is logically homogeneous.
The more involved case corresponds to the situation with
principally different kinds of motions or (and) hierarchy. It is
reasonable to define the second case as a complex resonance. In
several situations (i.e. stochastic resonance), when it is
impossible to define and compare clearly the essence of a state,
one considers a situation, when there is a sharp peak in the ratio
output-input at optimal value of external parameter.

An important moment should be mentioned regarding our concept. If
we consider some functional having different parameters,
functions, logical structures, and we optimize it over the entire
variables (besides some  fixed group of parameters or functions)
it could be stated that the essence of the whole system is the
same, as that one of a fixed group.

{\bf Anti-resonance.} Let me now analyze the resonance situation
with the opposite goal: to use the high levels of hierarchy to
achieve a maximal negative effect. This is a situation not too
rare in living systems.

We define anti-resonance as a situation, when:\\
1) a resonance is possible for some value of external parameter;\\
2) it is possible to define the opposite phase transformation of the parameter;\\
3) at the opposite phase values of the parameter  there is either\\
   a) an exponential damping of a motion, or\\
    b) a new feature (opposite in some sense  to those at the
resonant parameter case) arose in a resonant way.

The phenomenon is very complex. Thus, we are investigating the
simplest models, trying to reveal those  situations in complex
systems, when such phenomenon is possible.  Let us consider the
pendulum with $x(0)=x_0,x'(0)=0$, when the frequency varies with
some small amplitude h [54]:
\begin{eqnarray}
\label{e49} \frac{d^2 x}{d t^2}=-w^2(1+h\cos(2wt+\phi))x.
\end{eqnarray}
Here $h\ll 1$, $w$ is a frequency. Taking $\cos(2wt+\phi)=\sin (2
w t )$, we get an exponentially amplified solution solution:
\begin{eqnarray}
\label{e50} x(t)=\exp(\frac{hw}{4}t)[\cos(wt)]
\end{eqnarray}
Choosing $\cos(2wt+\phi)=-\sin (2wt)$, we have an exponential
damping
\begin{eqnarray}
\label{e51} x(t)=\exp(-\frac{hw}{4}t)[\cos(wt)]
\end{eqnarray}
 For the original amplitude
A the damping period $T$ is
\begin{eqnarray}
\label{e52} T\sim \frac{4\ln A}{hw}.
\end{eqnarray}
In this situation the picture is symmetric (both amplification and
damping are possible). The other   situation is possible with only
resonant damping (like domino effect).

{\bf Nishimori line.} One considers [50,51] $N$ spins $s_i$ with
interaction Hamiltonian
\begin{eqnarray}
\label{e53} H=-\sum_{i_1..i_p}j_{i_1..i_p}s_{i_1}s_{i_2}..s_{i_p}.
\end{eqnarray}
There is a p-spin interaction here, couplings $j_{i_1..i_p}$ are
random quenched variables $\pm 1$ with probability
$\frac{1+m_0}{2}$ for the values 1 and $\frac{1-m_0}{2}$ for the
values $-1$.  It is possible to write the following probability
distribution:
\begin{eqnarray}
\label{e54} P(j_{i_1\dots i_p
})=\frac{\exp(\beta_0\j_{i_1..i_p})}{2\cosh(\beta_0)}.
\end{eqnarray}
The parameter $\beta_0$ resembles an inverse temperature.
 Using the invariance of the
Hamiltonian under transformation
\begin{eqnarray}
\label{e55} s_i\to s_iv_i,\qquad j_{i_1..i_p}\to
j_{i_1..i_p}v_{i_1}v_{i_2}..v_{i_p},
\end{eqnarray}
in [50,51] has been calculated exact energy of the model at
$\beta_0=\beta$. At $\beta=\beta_0$ our system has the best
ferromagnetic properties in a sense that the number of up spins
$\sum_i\frac{<s_i>}{|<s_i>|}$ is maximal at Nishimori temperature
[51]. In opposite phase, we can take $\beta=-\beta_0$. While the
order parameter are different in ferromagnetic and
antiferromagnetic phases, free energy is the same in both models,
as  $Z(j,\beta)=Z(j,-\beta)$ for the Hamiltonian (\ref{e53}).
 For the odd values of $p$
one has an optimal properties for the configuration $s_i=-1$.
Thus, there is a trivial anti-resonance according to our
definition. For the even values of $p$ (i.e. $p=2$) and bonds on
the links of hypercube lattices in d -dimensional space, there is
an antiferromagnetic ordering (an anti-resonance situation).

{\bf Anti-resonance in complex systems.} A search of
anti-resonance in stochastic resonance [53] is a very interesting
issue. The resonance is certainly a complex one, when the
deterministic harmonic motion has the same period as the
transition by noise. To construct the anti-resonance is
problematic, as stochastic resonance has not a phase to reverse
the resonance situation. In [54]  the stochastic resonance
explanation for the crashes and bubbles in financial markets
(using the Ising spin model) has been considered. There is no
phase for the noise to be reversed in stochastic resonance,  but
the information for the agents can certainly be positive or
negative, thus moving the market from the border of two phases to
one side.

During the last decade, the idea of evolution  or development at
the edge of chaos [7-8], related to complex adaptive systems, was
very popular. What about anti-resonance aspect of the origin of
life? It is the case, at least, for the hyper cycle model by Eigen
and Schuster [55].  One tries to construct self-replicating system
overpassing simple problem,i.e. mutations, but, unavoidably,
parasite creatures appear. As a result, there is a chance to
consume all the information via those parasite creatures. We see,
in some sense, a resonance picture with a chance for
anti-resonance. The virus evolution is also often  near the error
threshold (mutation catastrophe)\cite{ei02}. At the top level of
life there is a phenomenon of apostasis, when the cell could be
killed by a simple command.

Our point of view is the following: even if complex systems are
walking at the edge of chaos and climbing the mountains of fitness
landscapes (in case of biological evolution),   it is often a walk
near the precipice. For the evolution it is not so dangerous, as
only the survival of the species is crucial. One should be much
more careful with a rare or single systems, like humanity.

{\bf Complexity parameters and stability of complex systems}.

What parameters could be applied to analyze the complex systems?
Besides the edge of chaos parameter, reasonable for the error
threshold like systems, we can use Nishimori temperature like
parameter. In principle Parisi's replica symmetry scheme also
could be considered as a complexity parameter. Parisi's replica
symmetry method works successfully for the mean field models,
giving exact solutions [54]. For the Hamiltonian $H(j,s_i)$ the
ensemble average via quenched disorder $j$ in $n$ replica
formalism brings  to Hamiltonian $\sum_{l=1}^nH(j,s^l_i)$ order
parameters like
$q_{\alpha,\beta}=\sum_is^{\alpha}_is^{\beta}_i/N$, where
$\alpha,\beta$ are replica indices. At the zero replica limit
those parameters can be expressed via replica symmetry breaking
parameters (given in our case by Eqs. (44),(45)). This scheme is a
single for the whole system. Therefore, for the stability of the
complex system it is important that those parameters coincide in
different sub-systems.

 Very important problem is to look for complexity parameters in
living matter. In case of proteins there is a notion of "design
temperature", similar to the Nishimori temperature(see review
[58]). Here Hamiltonian $H(j,s)$ is a function of $j_i$
(amino-acids type in a sequence) and $s_l$ (conformations). The
couplings $j$ have a distribution like the one in Eq.(54):
\begin{eqnarray}
\label{e54} P(j)\sim \exp(-H(j,s)\beta_d).
\end{eqnarray}
where $s$ is some ferromagnetic like "native" configuration.
Perhaps the methodology of Nishimori line  could be applied to the
protein case.  According to Nishimori [51] system with quenched
disorder reveal best ferromagnetic properties at Nishimori
temperature. It will be interesting to check this conjecture
(Nishimori proved for the case of discrete spins) for the
polymers.

Besides the design temperature, well defined for the all proteins,
there is also a notion of designability, which describes how
robust is the ground state against the mutations. As a rule, the
high designable proteins have also efficient low design
temperature, but could be an exceptions (high design temperature
with hogh design degree) [61].

In principle it is possible to construct equations for the protein
evolution [59] as well as genome growth [60]. One can derive some
degree of ferromagneticity for those cases as well.  It is
interesting to compare complexity parameters of those different
aspects of living matter.

\section{Conclusion}

 We have rigorously solved the error threshold for
 optimal codes using Random Energy Model, calculating
magnetization and finite size corrections to free energy. This
approach was applied in our previous works where many results of
Shannon information theory about optimal coding were derived.
 There is an alternative method (replica
approach with Nishimiori line), working well also in the case of
realistic Low-Density Parity Check Codes (LDPC) [61] (see review
[62]). REM approach could not be applied directly in the case of
finite block coding, but it is much simpler. Main results of
information theory were derived in REM approach about 6-9 years
before those, found through alternative method: error threshold
for finite rate of information transmission [19] versus [63],
reliability exponent [25] versus [64], data compression [23]
versus [64]. Multichannel coding was analyzed first in [24]. It is
very interesting to check the universality class of codes with
finite blocks length [61], optimal codes with finite number spin
interaction. Unfortunately, an alternative method of [63-64] could
not be applied here directly.

 Carefully
investigating error threshold phenomena in REM, we have found
several
 criteria of complexity: (33) and (35),(39),(44) and (46)
which could be applied for complex adaptive systems. In [4-5] has
already been suggested to consider the subdominant part of the
entropy as a measure of complexity. We have enlarged their idea,
suggesting to use a subdominant part of free energy as a measure
of complexity. It is more universal, than the bulk free energy,
and could be considered as the next step in the hierarchy
energy-free energy-subdominant term in free energy. This hierarchy
could be continued. Complexity appears on the third level, at some
higher levels the life could appear.
 We admit that our approach catches the qualitative
idea about edge of chaos: in the complex phase, the probabilities
of ordered and disordered motions are equal Eq.(44), and
complexity properties damp exponentially outside the error
threshold
 point, Eqs (35),(39),(45).
 We adduced arguments that, unlike to SOC or ordinary critical theories, HOT
design  belongs to error threshold universality class of
complexity.
 There are few classes of subdominant term behavior:
 zero, or
exponentially decreasing subdominant terms for Markov, and Hidden
Markov models [6]; logarithmic corrections for critical theories
[16]; cubic root corrections for Sherrington-Kirkpatrick model;
 square root
corrections for error threshold, long-range SG model [5] and,
maybe, language.
First a complexity class should be identified from the
 empirical data,  to model complex  phenomenon.
As percolation or SOC models belong to the   universality class
[16], it is improbable that they can describe financial markets.
Originally only SOC criticality has been identified with a
qualitative idea of "edge of chaos". But we see that error
threshold class is higher, than the SOC, and  this complexity
class is likely connected with alive-like systems [7]. We have
introduced also the concept of anti-resonance, a phenomenon,
perhaps, typical for the birth (and existence!) of life and for
advanced complex adaptive systems.

 We have suggested to investigate, at first, the main
features of complexity to identify the large universality classes.
What other characteristics could be used for the further
characterization of complex phenomenon? Perhaps the language of
the system with its grammar, or, in physical systems, the
existence of local gauge invariance. In case of REM, formulated as
a spin model, there is a local gauge invariance  (see
Eq.(\ref{e55})). There is local scale invariance for the models of
[16]. Therefore two theories could be connected, according to our
complexity analysis. It is really the case, as has been proved in
[64]. We hope that other applications of such analysis are
possible. The spin-glass phase and error threshold border in REM
reveal the advantage of subdominant free energy approach to
complexity compared with the subdominant entropy one. The latter,
if be used as a complexity measure, produces  lower classes ($\sim
O(1)$ instead of $\ln N$ or $\sqrt{N}$). We have used free energy
to define the complexity. In general, when direct statistical
mechanics formulation of the problem is impossible, one can use a
variable, describing a manageable amount of motion on macroscopic
level. The context of the problem can contribute greatly make  a
proper choice. For example, in Eigen model the equivalent of
energy is fitness (with a minus sign). Free energy is
automatically defined as a minus selective ability (mean fitness)
of a group of configurations.

In section 4 the idea of essence of the complex system state was
used for several times. In case of spin glasses, the real state of
the system is defined in the replica space, with some probability
followed by the projection to the zero replicas (in Parisi's
theory). In case of Hidden Markov Models  the state is not
directly observable again,as we get an information via
probabilistic process. In quantitative linguistics, an abstract
linear space has been applied to catch the meaning of the words
[66]. Perhaps, the first example is quantum mechanics: there is a
unitary evolution of the state  in Hilbert space, and during the
measurement we have some probabilistic results.
 In all
those examples the state of complex system is not formulated
directly via observable, but instead in some hidden abstract
space, where the interpretation of the system (its motion) is
rather simple one (the formulation of spin-glass statistical
physics in replica space is much easier, than the zero replica
limit, and formulation of Schrodinger equation is easier than
quantum theory of measurement).
 We assume that it is an important feature of complex
system: the real state of the system is in abstract hidden space,
and  can be observed in reality only in a probabilistic way.
Therefore we suggest a "principle of expanded pre-reality": to
solve complex problem one should reformulate the problem in some
internal,hidden, wider space ("pre-reality"), then return back to
the observable space ("reality") in a probabilistic way.

In lieu of the results, it is very important to look for
anti-resonance phenomenon in stochastic resonance. Unfortunately,
early attempts to find it have not been successful (H. Wio,
private communication). Another important problem is to identify
the universality class of turbulence. An accurate numerical
analysis to identify the universality class (Y. Sinai, private
communication) is likely possible for the case of Burgers
turbulence. According to the whole experience of complex systems
and our "pre-reality" principle, to succeed in turbulence solution
one should formulate the problem in a wider abstract space, then
return back to observable. It is very important to investigate the
language models [66], and latent semantic analysis [67] in our
approach. As we mentioned, the results of [31] (by means of
entropy analysis) already supports the idea that language belongs
to the error threshold class. The investigation of the semantic is
much deeper. The singular value decomposition in [66-67]
qualitatively resembles the fracturing of couplings into
ferromagnetic and noisy ones.

 {\bf Acknowledgments.} I am grateful
to C. Biebricher, S. Chen, V. Priezjev, Y. Sinai, D.Saad and H.
Wio for discussion, and anonymous referee for the valuable
remarks.
    Four years ago, when I began my studies,
I had a discussion with P.Bak in London. I greatly  appreciate his
 support of my idea to connect complex phenomena with error threshold. This work was partially
supported by the National Science Council of
  Taiwan under grant No. NSC 91-2112-M-001-056.


\begin{thebibliography}{99}
\bibitem{zu90} C.H. Bennet, p.215 in {\it Complexity,Entropy  and the physics of information},W.H. Zurek
 editor,Addison-Wesley,1990.
\bibitem{ge94} M. Gell-Mann, p.17 in {\it Complexity:Metaphors, Models and Reality}, Eds G.Cowan, D. Pines,
D. Meltzer,SFI-XIX,1994.
\bibitem{gl04} M. Gell-Mann, S. Lloyd, p.387 in {\it Nonextensive Entropy:
Interdisciplinary Applications},
  M. Gell-Mann and C. Tsallis, Eds., Oxford University Press, New York,
  (2004).
\bibitem{gr86} P. Grassberger, Int. J. Theor.Phys. {\bf 25},907(1986)
\bibitem{bi01} W.Bialek,I. Nemenman, and N. Tishby, Neural Computation {\bf
13},2409(2001).
\bibitem{cr02} J.P.Crutchfield,D.P.Feldman, cond-mat/0102181.
\bibitem{ka93} S.A.Kauffmann, {\it The Origns of Order, Self-Organization and
Selection in Evolution},Oxford Uiversity Press,NY,1993
\bibitem{ba87}P.Bak,C.Tang,K.Wisenfield,Phys.Rev.Lett. {\bf
59},381(1987).
\bibitem{b00}P. Bak, private communication, London 2000
\bibitem{dh87}D.Dhar,Phys.Rev.Lett. {\bf 67},1713(1987).
\bibitem{chi85}I.Chisar,J.Korner,{\it Information Theory},Moscow,1985
\bibitem{sh99}J.S. Shiner and M. Davison, Phys. Rev. E. {\bf 59},
1459(1999)
\bibitem{cr99}J. P. Crutchfield and C. R. Shalizi, Phys. Rev. E {\bf 59}275(1999)
\bibitem{ja82}Jayenes, E.T. {\it Papers on Probability, Statistics and Statistical Physics},
editor R.D. Rosenkrantz,Reydel,Dordrecht,1982.
\bibitem{b20}E. J. Chaisson, Cosmic evolution, Harvard Univ. Press, London,2001
\bibitem{ca87}J.L.Cardy,I.Peschel,
Nucl. Phys. {\bf B300,FS22},377(1988).
\bibitem{so89} N.Sourlas, Nature {\bf 239},693(1989) .
\bibitem{de80} B.Derrida, Phys. Rev. B {\bf 24},2613(1981).
\bibitem{sa1}D.B.Saakian,JETP lett.{\bf 55},n.2(1992).
\bibitem{ru93}P.Rujan, Phys.Rev.Lett. {\bf 70},2968(1993).
\bibitem{sa2} D.B.Saakian, JETP Lett.,{\bf 61},n.8(1995).
\bibitem{sa3} A.E.Allahverdyan,D.B.Saakian,  Teor. Math. Fizika {\bf
109},1574(1996).
\bibitem{sa4} D.B.Saakian, Teor.Mat.Fizika,{\bf 97},1199(1993).
\bibitem{sa5}A.E. Allakhverdyan, D.B.Saakian,  JETP {\bf
111},n.3(1997).
\bibitem{sa6}A.Allakhverdian,D.Saakian, Nucl. Phys. {\bf
B498},604(1997).
\bibitem{ks04}Y.~Kabashima and D.~Saad, J.Phys. {\bf
37A},R1(2004).
 \bibitem{ei71}  Eigen,M.  {\it Naturwissenschaften}  {\bf 58}, 465(1971).
 \bibitem{bbw97} Baake,E.,Baake,M.\& Wagner,H.  Phys.Rev.Lett. {\bf 78}, 559(1997)
\bibitem{sh04}Saakian,D.B.\& Hu,C.K.
{\it Phys.Rev.E} {\bf 69},021913(2004)
\bibitem{sh04a}Saakian, D.B.\& Hu,C.K., Phys.Rev.E,{\bf 69},046121(2004)
\bibitem{hi90} Hilberg,W. Frequenz {\bf 44},243(1990).
\bibitem{tsa88} Tsallis,C. J.Stat.Phys.{\bf 52},479(1988).
\bibitem{pe02}Costa M., Goldberger A. Peng C-K,Phys.Rev.Lett. {\bf 89},068102(2002)
\bibitem{in97}R.S. Ingarden,A.Kossakowski,M.Ohya, {\it Information dynamics and open systems},
Kluber,1997.

\bibitem{pe01}Ch.Renner,J.Peinke,R.Friedrich,cond-mat/0102494
\bibitem{so01} D. Sornette,Physica {\bf A290},211(2001).
\bibitem{lu99}T. Lux, M. Marchesi, Nature {\bf 297},498(1999)
\bibitem{do00}J. Doyle, J.M. Carlson,Phys. Rev. Lett {\bf
84},2529(2000).
\bibitem{do00a}J. Doyle, J.M. Carlson,Phys. Rev. Lett {\bf
84},5656(2000).
\bibitem{fa02} M.E.Newman,M.Girvan,J.D.Farmer,PRL,{\bf
89},028301(2002).
\bibitem{co91}T.M.Cover and J.A. Thomas,{\it Elements of Information
Theory}, ,Wiley,NY, 1991.

\bibitem{ko01}T.Koski,{\it  Hidden markov models for
bioinformatics},Kluber Academic Publishers, Dordrecht, 2001.
\bibitem{new04}J.Park and M.E.J.Newman,cond-mat 0405566.
 \bibitem{eigen89} M. Eigen, J. McCaskill,P. Schuster,
   Adv.  Chem. Phys.    {\bf 75}, 149 (1989).
\bibitem{pe03} A.S. Perelson, Review of Modern
Physics, v.69,1219(1997).

\bibitem{sk75}D.Sherrington,S.Kirkpatrick, Phys.Rev.Lett. {\bf
35},1792(1975)
\bibitem{cr04}
A. Crisanti,L. Leuzzi,G. Parisi,T.Rizzo, Phys. Rev. Lett. {\bf
92},127203(2004)
\bibitem{b28}G.Parisi,F.Ritort,F. Slanina,JPA,{\bf 26}(1993)3775
\bibitem{wi02}A. T. Winfree, Science {\bf 298},2336(2002).
\bibitem{ni81}H.Nishimori,Prog. Theor. Phys. {\bf 66},1169(1981).
\bibitem{ni02}H. Nishimori, Physica A {\bf 306}, 68(2002).
\bibitem{ll}L.D. Landaw,E.M. Lifshitz, Classical mechanics.
\bibitem{be81}Benzi R. ,Sutera S. and Vulpiani A. , J. Phys. {\bf
A14},L453(1981).
\bibitem{kh03}A.Krawiecki,J.A. Holyst,Physica {\bf
A317},597(2003).
\bibitem{ei79}M.Eigen, P.Schuster,{\it  The hypercycle-A principle of Natural
self-Organization},Springer,Heidelberg,1979.
\bibitem{ei02}M.Eigen,PNAS,{\bf 99},13374(2002).
\bibitem{pa79}G.Parisi,PRL,{\bf 43},1754(1979).
\bibitem{gr00}V.S. Pande, A.Yu.Grossberg and T.
Tanaka,Rev.Modern.Phys.{\bf 72},260(2000).
\bibitem{ta96}R.Melin, H.Li,N.S. Wingreen and C. Tang,  Journal of Che,{\bf
273},666(1996).
\bibitem{lee03}L.-C. Hsieh, L. Luo, F. Ji, and H.
C. Lee,PRL,{\bf 90},018101(2003).
\bibitem{dee04}D.J. Earl and M.W. Deem, PNAS, (2004).
\bibitem{ma95}D. MacKay, R. Neal, Lecture notes in computer
science, {\bf 1025},p.100,Berlin,Springer,1995.
\bibitem{b30}Y.Kabashima,D.Saad,Euro.phys.lett{\bf 45},97(1999).
\bibitem{sa03}N. Skantzos,J.Mourik,D.Saad, J.Phys. {\bf
36A},11131(2003.)
\bibitem{mu02}T.Murayama,J.Phys.{\bf A35}L95(2002).
\bibitem{b17} D. Saakian, Phys.Rev PRevE,{\bf  65}(2002) 67104
\bibitem{ac04}D. Aerts,M.Czachor,J.Phys.A. {\bf 37},L123(2004)
\bibitem{la97}T.K. Landauer, k.P. Foltz and D.Laham,Discourse
Process,{\bf 25},259(1998).
\end{thebibliography}
\end{document}